\documentclass[sigplan,screen,nonacm]{acmart}

\usepackage{multirow}
\usepackage[capitalise,noabbrev]{cleveref}
\usepackage{caption}
\usepackage{subcaption}
\usepackage{listings}
\lstdefinelanguage{futhark}
{
  morekeywords={
    do,
    else,
    for,
    if,
    in,
    include,
    let,
    loop,
    then,
    type,
    val,
    while,
    with,
    module,
    def,
    entry,
    local,
    open,
    import,
    assert,
    match,
    case,
  },
  sensitive=true, % Keywords are case sensitive.
  morecomment=[l]{--}, % l is for line comment.
  morestring=[b]" % Strings are enclosed in double quotes.
}

\newcommand{\id}[1]{\ensuremath{\mathit{#1}}}
\newcommand{\kw}[1]{\ensuremath{\mathtt{#1}}}
\newcommand{\Let}{\kw{let}}
\newcommand{\In}{\kw{in}}
\newcommand{\If}{\kw{if}}
\newcommand{\Then}{\kw{then}}
\newcommand{\Else}{\kw{else}}
\newcommand{\Filter}{\kw{filter}}
\newcommand{\Or}{\kw{or}}
\newcommand{\Groupby}{\kw{groupby}}
\newcommand{\Partition}{\kw{partition}}
\newcommand{\Scatter}{\kw{scatter}}
\newcommand{\Hist}{\kw{hist}}
\newcommand{\Map}{\kw{map}}
\newcommand{\Sum}{\kw{sum}}
\newcommand{\Mod}{\kw{mod}}
\newcommand{\Fst}{\kw{fst}}
\newcommand{\Snd}{\kw{snd}}
\newcommand{\Presum}{\kw{presum}}

\newcommand{\Sort}{\kw{sort}}
\newcommand{\Segor}{\kw{segor}}
\newcommand{\Hash}{\kw{hash}}
\newcommand{\Random}{\kw{random}}
\newcommand{\Iota}{\kw{iota}}
\newcommand{\Unzip}{\kw{unzip}}
\newcommand{\Unit}{\mathbf{unit}}
\newcommand{\Int}{\mathbf{int}}
\newcommand{\Bool}{\mathbf{bool}}

\newcommand{\Rep}{\kw{rep}}
\newcommand{\Zip}{\kw{zip}}

\DeclareMathOperator{\concat}{+\kern -0.2em+}
\begin{document}

\title{Towards Efficient Hash Maps in Functional Array Languages}

\author{William Henrich Due}
\email{widu@di.ku.dk}
\orcid{0009-0003-8195-4462}
\affiliation{%
  \institution{University of Copenhagen}
  \city{Copenhagen}
  \country{Denmark}
}

\author{Martin Elsman}
\email{mael@di.ku.dk}
\orcid{0000-0002-6061-5993}
\affiliation{
  \institution{University of Copenhagen}
  \city{Copenhagen}
  \country{Denmark}
}

\author{Troels Henriksen}
\email{athas@sigkill.dk}
\orcid{0000-0002-1195-9722}
\affiliation{%
  \institution{University of Copenhagen}
  \city{Copenhagen}
  \country{Denmark}
}

\begin{abstract}
  We present a systematic derivation of a data-parallel implementation of
  two-level, static and collision-free hash maps, by giving a functional
  formulation of the Fredman et al. construction, and then flattening it. We
  discuss the challenges of providing a flexible, polymorphic, and abstract
  interface to hash maps in a functional array language, with particular
  attention paid to the problem of dynamically sized keys, which we address by
  associating each hash map with an arbitrary context. The algorithm is
  implemented in Futhark, and the achieved GPU execution performance is compared
  on simple benchmark problems. We find that our hash maps outperform
  conventional tree/search-based approaches. Furthermore, our implementation is
  compared against the state-of-the-art cuCollections library, which is
  significantly faster for hash map construction, and to a lesser degree for
  lookups. We explain to which extent the performance difference is due to
  low-level code generation limitation in the Futhark compiler, and to which
  extent it can be attributed to the data-parallel programming vocabulary not
  providing the constructs necessary to express the equivalent of the algorithms
  used by cuCollections. We end by reflecting to which extent the functional
  array language programming model could, or should, be extended to address
  these weaknesses.
\end{abstract}


\maketitle

\section{Introduction}
Hash maps are a commonly used data structure for storing key-value pairs.
They are also well suited for data-parallel programming just like arrays.
As an example you can map over its values or reduce its values to a single
value.
Despite their usefulness, hash maps are not commonly used in functional array
languages.
They are often used to implement certain constructs in APL dialects, but in
those cases the hash maps are part of the interpreter (typically written in C),
and not implemented in the language itself.
In this paper we describe how to implement a hash map in a functional array
language in a way that is asymptotically work efficient, and also performs
reasonably in practice.
We describe how such a hash map can be derived from a functional
algorithm. Concretely, we demonstrate how to implement the two-level hash set
construction by Fredman et al. in a functional array language.
Furthermore, we demonstrate how to interact with hash maps using an SML style
module system, in such a way that it is possible to use the hash maps with
irregular datatypes as keys.

Our specific contributions are as follows:

\begin{itemize}
\item We demonstrate how the two-level hash map construction of Fredman et
  al~\cite{10.1145/828.1884} can be expressed in a data-parallel vocabulary,
  which requires nontrivial segmented operations.
\item We show how hash maps can be exposed in a modular and generic way in a
  data-parallel language, using an ML-style module system, including how to
  handle complicated key types such as strings.
\item We perform a performance comparison demonstrating that data-parallel hash
  maps running on GPUs outperform data structures based on binary search on both
  integer and string keys, but that the achieved performance still falls
  short of what is possible in CUDA.
\end{itemize}

\subsection{Outline}  % maybe delete subsection title
This paper is structured as follows:
\begin{itemize}
  \item \Cref{sec:two-level-hashmaps} describes a functional algorithm for the two-level hash set construction by Fredman et al. and how it can be flattened
  to a data-parallel algorithm.
  The description will use SML like pseudo code
  to describe the algorithm with common array operations, which can be
  parallelised.
  \item \Cref{sec:interface} describes an interface for hash maps
  suitable for a functional array language.
  Interfaces are written in Futhark's ML-style module system.
  \item \Cref{sec:benchmarks} describes the benchmarks used to compare the performance of our hash map implementation against existing alternatives in
  array languages written in Futhark and the cuCollections library.
\end{itemize}

\section{Two-Level Hash Maps}
\label{sec:two-level-hashmaps}
There are many ways of implementing a hash map, but only a few are suitable for
a data-parallel functional array language.
A problem that has to be solved during construction of a hash map is accounting
for collisions.
One could imagine using a chained hash map where we have two key-value pairs to
be inserted at the same index in the hash map in parallel due to a hash
collision.
In a language like CUDA we could use atomic operations to insert them into the
hash map in some order depending on hardware specifics.
In a functional setting such a strategy is problematic, because it requires
nondeterministic and impure operations.
One of the possible solutions could be sorting the keys by their hashes and then
inserting them in order into the hash map.
Strategies based on sorting, however, are inefficient and should be avoided as
the performance will drop to the performance of binary search trees or Eytzinger
trees \cite{DBLP:journals/corr/KhuongM15}.

A solution to the collision problem is to use the two-level hash set
construction by Fredman et al. \cite{10.1145/828.1884} (FKS).
This construction has a hash function that is collision-free such that we do
not have to consider the insertion order of the keys.
To use the algorithm efficiently on real world hardware, we need the ability to
compute histograms \cite{10.5555/3433701.3433830} using atomic operations.
These are used to count the occurrences of a given hash in the hash set. An
alternative is to use sorting and a segmented scan, but sorting is once again
expensive.

\subsection{Prerequisites}
\label{seq:prerequisites}
This article assumes a basic familiarity with data-parallel programming and
functional programming. We will use a data-parallel vocabulary that is found in
\cref{fig:vocabulary} to describe the algorithm.
\begin{figure}

\begin{align*}
  \Fst & : (\alpha, \beta) \to \alpha \\
  \Snd & : (\alpha, \beta) \to \beta \\
  \Map & : (\alpha \to \beta) \to [n]\alpha \to [n]\beta \\
  \Iota & : \Int \to [n]\Int \\
  \Zip & : [n]\alpha \to [n]\beta \to [n](\alpha, \beta) \\
  \Unzip & : [n](\alpha, \beta) \to ([n]\alpha, [n]\beta) \\
  \Partition & : (\alpha \to \Bool) \to [n]\alpha \to ([m]\alpha, [k]\alpha) \\
  \Filter & : (\alpha \to \Bool) \to [n]\alpha \to [m]\alpha \\
  \Presum & : [n]\Int \to [n]\Int \\
  \Sum & : [n]\Int \to \Int \\
  \Or & : [n]\Bool \to \Bool \\
  \Segor & : [m]\Int \to [n]\Bool \to [n]\Bool \\
  \Rep & : \Int \to \alpha \to [n]\alpha \\
  \Scatter & : [m]\alpha \to [n]\Int \to [n]\alpha \to [m]\alpha \\
  \Sort & : [n](\Int, \alpha) \to [n]\alpha \\
  \Groupby & : (m: \Int) \to [n]\Int \to [n]\alpha \to [m][]\alpha \\
  \Random & : \Unit \to [c]\Int \\
  \Hash & : [c]\Int \to \alpha \to \Int \\
  \Hist & : (n: \Int) \to [m]\Int \to [m]\Int \to [n]\Int \\
  \mathtt{repiota} & : [n]\Int \to [m]\Int
\end{align*}

\caption{The data-parallel vocabulary used for explaining the construction of
  two-level hash maps.}
  \label{fig:vocabulary}
\end{figure}

The size of the arrays is denoted by $n$, $m$, and $k$ and is used to denote
that the arrays have different sizes for a given function while $c$ is a
constant size.
The first projection of tuples is denoted by $\Fst$ and the second by $\Snd$.
Applying a function on every element in an array is denoted by $\Map$.
Producing an array of integers from $0$ to $n - 1$ is done by $\Iota \: n$.
Constructing an array of tuples from two arrays is done by $\Zip$ and $\Unzip$
is the inverse to $\lambda (a, b) \to \Zip \: a \: b$.
For partitioning an array into two arrays based on a predicate we use
$\Partition$, the array in the first position of the result is for elements
where the predicate holds true and the second is for elements where it holds
false.
Filtering every element in an array that does not satisfy a predicate we use
$\Filter$.
The operation for computing a prefix sum of an array of integers is denoted by
$\Presum$.
The operation for computing the sum of an array of integers is denoted by
$\Sum$.
The operation for computing the logical or over an array of booleans is $\Or$.
A segmented scan using the logical or operation can be computed using $\Segor$
(i.e., a segmented-$\Or$).
An expression $\Rep \: n \: a$ results in an array where the element $a$ is
replicated $n$ times.
The $\Scatter$ function takes an array to be written to, an array of indices,
and an array of values to write at the given indices.
Sorting is denoted by $\Sort$, which sorts the array by the first element and
discards it (e.g., $\Sort \: [(2, x), (1, y)] = [y, x]$).
To group an array of elements by an array indices is denoted by $\Groupby$,
which produces an irregular array of arrays (e.g., $\Groupby \: 4 \: [2, 0, 2]
\: [x, y, z] = [[y], [], [x, z], []]$).
Producing an array of random integers of some fixed size $c$ is denoted by
$\Random$. This function must have the property that returned values form a distribution such that when given as
constants to $\Hash$ it forms a universal family of hash functions
\cite{CARTER1979143} \cite[p.~2]{DBLP:journals/corr/Thorup15}.
We treat the $\Random$ function as impure, but it is readily implemented in a
functional language through conventional state passing mechanism.
Computing histograms are denoted by $\Hist \: n \: is \: vs$, here $n$ is the
number of bins (initialized to 0), $is$ is what bucket a given value in $vs$
belongs to (i.e., like $\Scatter$ but allows for duplicate indices where their
values are summed).
A replicated iota \cite{Elsman:2019:DFE:3315454.3329955} is denoted by
$\mathtt{repiota}$, which is like $\Iota$ but where each integer in the argument
array determines the number of times an element is repeated (e.g.,
$\mathtt{repiota} ~[2, 3, 1] = [0, 0, 1, 1, 1, 2]$).

\subsection{The Construction}
A \emph{hash set construction} for a non-empty finite set of keys $K$ (taken
from a universe $U$), with cardinality $|K|=n$, and parameterised over a
constant $c$, is a tuple
$$(\id{const}, \id{consts}) : ([c]\Int, [n][c]\Int)$$
which defines a hash function $\mathtt{h}$ as follows:
\begin{align*}
  & \mathtt{h} \: k = \id{offsets}[\kw{g} \: k] + (\Hash \: \id{consts}[\kw{g} \: k] \: k \: \Mod \: \id{shape}[\mathtt{g} \: k]) & & \\
  & \kw{g} \: k = (\Hash \: \id{const} \: k) \: \Mod \: n \\
  & \id{shape} = (\Map \: (\lambda s \to s^2) \circ \Hist \: \id{n} \: (\Map \: g \: K)) \: (\Rep \: n \: 1) \\
  & \id{offsets} = \Presum \: \id{shape}
\end{align*}

We further say that such a hash set construction is \emph{well-formed} and
\emph{collision-free} if the following properties hold:
\begin{enumerate}
\item $\{\kw{h}~k ~|~ k\in K\} \subseteq \{0,\ldots,s-1\}$, where $s = \Sum \: \id{shape}$
\item $|\{\kw{h} \: k ~|~ k \in K\}| = |K|$
\end{enumerate}

The goal now is, given a non-empty finite set of keys, to find a collision-free
and well-formed hash set construction.
Given such a construction, it is then possible to create an array $\id{arr}$ of size $s$ and place every key $k$
at index $\kw{h}~k$ and fill remaining indices with any key $k \in K$.
To assert if for some $a \in U$ (the universe), we also have $a \in K$, we
simply test if $\id{arr}[\kw{min}~(s - 1) \: (\kw{h}~a)] = a$ holds true.
Such a hash set can be extended to a hash map by storing the value in an
accompanying array where $\kw{h}~a$ is used for indexing.

\subsection{Functional Construction}
The data-parallel variant of the two-level hash set construction can be derived
from a functional variant.
This variant comes from an formulation of the two-level hash set construction
described in the FKS paper.
\subsubsection{Level One}
\begin{align*}
  & \mathbf{def} \: \kw{make}_1 \: (\id{keys}: [n]\alpha) = \\
  & \quad \Let \: \id{const} = \Random \: () \\
  & \quad \Let \: \id{hashes} = \Map \: ((\Mod \: n) \circ \Hash \: \id{const}) \: \id{keys} \\
  & \quad \Let \: \id{shape} = \Hist \: \id{n} \: \id{hashes} \: (\Rep \: n \: 1) \\
  & \quad \Let \: \id{consts} = (\Map \: \kw{make}_2 \circ \Groupby \: n \: \id{hashes}) \: \id{keys} \\
  & \quad \In \: (\id{const}, \id{consts})
\end{align*}
The first step of the FKS construction is to choose some random constants for a
hash function.
This is used to compute the \textit{hashes} of each keys.
Afterwards every key is grouped by its hash and then the $\kw{make}_2$
function is called on each subarray of keys such that a collision-free hash set
for every subarray can be found.
We consider every hash set made by $\kw{make}_2$ to be the second level of the
hash set.

\subsubsection{Level Two}
The construction of the second level hash set needs the following functions.
\begin{align*}
  & \mathbf{def} \: \kw{hashes} \: (\id{const}: [c]\Int) \: (\id{keys}: [m]\alpha) = \\
  & \quad \Map \: ((\Mod \: m^2) \circ \Hash \: \id{const}) \: \id{keys}
\end{align*}
A function which computes the hashes of the keys.
\begin{align*}
  & \mathbf{def} \: \kw{collision} \: (\id{hashes}: [m]\Int) = \\
  & \quad (\Or \circ \Map \: (> 1) \circ \Hist \: m^2 \: \id{hashes}) \: (\Rep \: m \: 1)
\end{align*}
A function which checks if the constants leads to a collision-free hash
function.
\begin{align*}
  & \mathbf{def} \: \kw{make}_2 \: (keys: [m]\alpha) = \\
  & \quad \Let \: \id{cs} = \Random \: () \\
  & \quad \Let \: \id{hs} = \kw{hashes} \: \id{cs} \: \id{keys} \\
  & \quad \In \: \If \: \kw{collision} \: \id{hs} \: \Then \: \kw{make}_2 \: \id{keys} \: \Else \: \id{cs}
\end{align*}
The function $\kw{make}_2$ continues to generate constants $\id{cs}$ for a hash
function until it finds constants that leads to a collision-free hash function.

The problem now is $\kw{make}_2$ is mapped over an irregular array of keys.
This is not data-parallel but task-parallel, but it is possible to use
flattening to get a data-parallel version of the algorithm.
\subsubsection{Analysis}
Almost every line of $\kw{make}_1$ does $O(n)$ work except for the mapping of
$\kw{make}_2$ over the grouped keys which creates $\id{consts}$.
Grouping can be done in $O(n)$ work and $\kw{make}_2$ does $O(n_i^2)$ work where
$n_i$ is the size of the subarray.
The sum of every subarray of size $n_i^2$ has the expected asymptotic value
of $O(n)$ and for any application of $\kw{make}_2$ on a subarray it is
expected to do at most $2$ trials to find a collision free hash function (i.e.,
expected $O(1)$ recursive calls) \cite{abrahamsen2020static}.
The total work done by mapping $\kw{make}_2$ over every subarray has the
expected work of $O(n)$ and so does the total work done by constructing the
hash set.

\subsection{Flattened Construction}
First $\kw{make}_1$ has to be changed such that a flattened version of
$\kw{make}_2$ can be used.
\begin{align*}
  & \mathbf{def} \: \kw{make}_1 \: (\id{keys}: [n]\alpha) = \\
  & \quad \Let \: \id{const} = \Random \: () \\
  & \quad \Let \: \id{hashes} = \Map \: ((\Mod \: n) \circ \Hash \: \id{const}) \: \id{keys} \\
  & \quad \Let \: \id{skeys} = (\Sort \: \circ \: \Zip \: \id{hashes}) \: \id{keys} \\
  & \quad \Let \: \id{shape} = \Hist \: n \: \id{hashes} \: (\Rep \: n \: 1) \\
  & \quad \Let \: \id{consts} = \kw{segmake}_2 \: \id{shape} \: \id{skeys} \\
  & \quad \In \: (\id{const}, \id{consts})
\end{align*}
We have to group the keys by their hashes in a flattened manner.
The flattened equivalent of $\Groupby$ is to sort the keys by their hashes using
a stable sorting algorithm.
However, this does not account for empty arrays created by $\Groupby$, these are
going to be handled later in $\kw{segmake}_2$ (The lifted version of
$\kw{make}_2$).
Also as a step in the derivation we sort the keys by their hashes, but in the
end we will not need sorting.

The $\id{shape}$ is computed in the same manner as before, but it does not
reflect the flattened version of the grouped keys that is \id{skeys} due to
empty subarrays are not account for.
This is handled in $\kw{segmake}_2$, so the shape and the sorted keys can be
passed along to the lifted version of $\kw{make}_2$ as is.

To lift $\kw{make}_2$ the following helper function will be lifted $\Random$,
\texttt{hashes}, and \texttt{collision} such that they use the shape to map over
the flattened array. For these helper functions we will continue to not consider
the empty subarrays.
\begin{align*}
  & \mathbf{def} \: \kw{segrandom} \: (\id{shape}: [m]\Int) = \\
  & \quad (\Map \: \Random \circ \Rep \: m) \: ()
\end{align*}
Each subarray needs a constant for its hash function, so the lifted version of
$\Random$ is to generate a constant for every subarray.
\begin{align*}
  & \mathbf{def} \: \kw{seghashes} \: (\id{shape}: [m]\Int) \: (\id{consts}: [m][c]\Int) \\
  & \qquad (\id{keys}: [n]\alpha) = \\
  & \quad \Let \: \id{offsets} = (\Presum \circ \Map (\lambda s \to s^2)) \: \id{shape} \\
  & \quad \Let \: \id{ii}_2 = \kw{repiota} \: \id{shape} \\
  & \quad \In \: \Map \: (\lambda i \to \\
  & \quad \quad \Let \: j = \id{ii}_2[i] \\
  & \quad \quad \Let \: (\id{o}, \id{cs}, \id{s}) = (\id{offsets}[j], \id{consts}[j], \id{shape}[j]) \\
  & \quad \quad \In \: \id{o} + (\Hash \: \id{cs} \: \id{keys}[i]) \: \Mod \: s^2) \: (\Iota \: n)
\end{align*}
To lift \texttt{hashes} we need to know the offsets of the flattened array of
squared subarray sizes.
This is done by squaring every element of the shape and then taking the prefix-sum.
$\id{ii}_2$ is an array of indices, it associates a given key in \textit{keys} with
what subarray it belongs to given by the \textit{shape}.
Lastly the hashes can be computed by adding the offset to the hash given by the
subarrays hash function.
\begin{align*}
  & \mathbf{def} \: \kw{segcollisions} \: (\id{shape}: [m]\Int) \: (\id{hashes}: [n]\Int) = \\
   & \quad \Let \: \id{sshape} = \Map (\lambda s \to s^2) \: \id{shape} \\
  & \quad \Let \: \id{flat} = \Sum \: \id{shape} \\
  & \quad \Let \: \id{counts} = \Hist \: \id{flat} \: \id{hashes} \: (\Rep \: n \: 1) \\
  & \quad \In \: (\Segor \: \id{sshape} \circ \Map \: (> 1)) \: \id{counts}
\end{align*}
The lifted version of the $\kw{collision}$ function can now be easily
derived since we can find the flat size of $\id{keys}$ by squaring every
element of shape and summing it.
Then since now the $\id{hashes}$ are indices into the flat array it is
simply possible to compute a histogram to count the number of collisions.
Afterwards use a segmented-or to check if any of the subarrays have a collision.

The next problem is how to translate the last if expression of $\kw{make}_2$
into a lifted version.
The goal is to stop working on all subarrays where the hash function leads to
zero collisions and then return its constants.
The problem here is the subarrays may finish out of order.
To solve this, every subarray size in shape and resulting constant for a hash
function is associated with an index.
This index can then be used to reorder the constants at the end.
For this part a lifted version of the results and new keys that will be passed
along to the recursive call will be given.
\begin{align*}
  & \mathbf{def} \: \kw{segresult} \: (\id{ishape}: [m](\Int, \Int)) \: (\id{consts}: [m][c]\Int) \\
  & \qquad (\id{collisions} : [m]\Bool) = \\
  & \quad \Let \: (\id{is}, \id{shape}) = \Unzip \: \id{ishape} \\
  & \quad \Let \: \id{iconsts} = \Zip \: \id{is} \: \id{consts} \\
  & \quad \Let \: (\id{notdone}, \id{done}) = \\
  & \qquad (\Partition \: (\lambda i \to \id{collisions}[i])) \: (\Iota \: m) \\
  & \quad \Let \: \id{iconsts'} = \Map \: (\lambda i \to \id{iconsts}[i]) \: \id{done} \\
  & \quad \Let \: \id{ishape'} = \Map \: (\lambda i \to \id{ishape}[i]) \: \id{notdone} \\
  & \quad \In \: (\id{ishape'}, \id{iconsts'})
\end{align*}
The first step is to associate \textit{consts} with the index of the subarray it
belongs to.
Then the subarray indices are partitioned into two arrays based on if it has no
collisions and is therefore done.
Lastly these indices can be used to produce an array of the constants which are
done and can be returned and the remaining subarrays shape which needs to be
worked.
\begin{align*}
  & \mathbf{def} \: \kw{segkeys} \: (\id{shape}: [m]\Int) \: (\id{collisions}: [m]\Bool) \\
  & \qquad (\id{keys}: [n]\alpha) = \\
  & \quad \Let \: \id{ii}_2 = \kw{repiota} \: \id{shape} \\
  & \quad \Let \: \id{colkeys} = \Map \: (\lambda i \to \id{collisions}[i]) \: \id{ii}_2 \\
  & \quad \In \: (\Map \: \Snd \circ \Filter \: \Fst \circ \Zip \: \id{colkeys}) \: \id{keys}
\end{align*}
To find the keys that will be worked on in a recursive call the $ii_2$ array is
created and every key that lead to no collisions can simply be removed.
The remaining keys are member of a subarray that lead to collisions.
\begin{align*}
  & \mathbf{def} \: \kw{segmake}'_2 \: (\id{ishape}: [m](\Int, \Int)) \: (\id{keys}: [n]\alpha) = \\
  & \quad \If \: n = 0 \: \Then \: [] \: \Else \\
  & \quad \Let \: \id{shape} = \Map \: \Snd \: \id{ishape} \\
  & \quad \Let \: \id{consts} = \kw{segrandom} \: \id{shape} \\
  & \quad \Let \: \id{hashes} = \kw{seghashes} \: \id{shape} \: \id{consts} \: \id{keys} \\
  & \quad \Let \: \id{collisions} = \kw{segcollisions} \: \id{shape} \: \id{hashes} \\
  & \quad \Let \: (\id{ishape'}, \id{iconsts'}) = \\
  & \qquad \kw{segresult} \: \id{ishape} \: \id{consts} \: \id{collisions} \\
  & \quad \Let \: \id{keys'} = \kw{segkeys} \: \id{shape} \: \id{collisions} \: \id{keys} \\
  & \quad \In \: \id{iconsts'} \concat \kw{auxiliary} \: \id{keys'} \: \id{ishape'}
\end{align*}
Everything can now be put together in a function which takes the shape with its
associated indices and keys.
Then it uses every lifted function and adds \textit{iconsts'} to the result
and works on the remaining keys and shapes.
\begin{align*}
  & \mathbf{def} \: \kw{segmake_2} \: (\id{shape}: [n]\Int) \: (\id{keys}: [n]\alpha) = \\
  & \quad \Let \: \id{ishape} = \\
  & \qquad (\Filter \: ((\neq 0) \circ \Snd) \circ \Zip \: (\Iota \: n)) \: \id{shape} \\
  & \quad \Let \: (\id{is}, \id{consts}) = (\Unzip \circ \kw{segmake}'_2 \: \id{ishape}) \: \id{skeys} \\
  & \quad \In \: \Scatter \: (\Rep \: n \: (\Random \:  ())) \: \id{is} \: \id{consts}
\end{align*}
Lastly in the lifted version of $\kw{make}_2$ the shape is filtered for
empty subarrays together with their indices.
Then $\kw{segmake}'_2$ can be used to compute the constants which are then
scattered into an array such they have the correct ordering.
\subsection{Sortless Construction}
The last problem is that the sorting should not be used since it is inefficient.
\begin{align*}
  & \mathbf{def} \: \kw{make}_1 \: (\id{keys}: [n]\alpha) = \\
  & \quad \Let \: \id{const} = \Random \: () \\
  & \quad \Let \: \id{hashes} = \Map \: ((\Mod \: n) \circ \Hash \: \id{const}) \: \id{keys} \\
  & \quad \Let \: \id{shape} = \Hist \: n \: \id{hashes} \: (\Rep \: n \: 1) \\
  & \quad \Let \: \id{consts} = \kw{segmake}_2 \: \id{shape} \: \id{keys} \: \id{hashes} \\
  & \quad \In \: (\id{const}, \id{consts})
\end{align*}
To start off we simply remove the sorting and pass along the hashes and keys to
$\kw{segmake}_2$.
\begin{align*}
  & \mathbf{def} \: \kw{segmake_2} \: (\id{shape}: [n]\Int) \: (\id{keys}: [n]\alpha) \\
  & \qquad  (\id{hashes}: [n]\Int) = \\
  & \quad \Let \: \id{offsets} = (\Presum \circ \Map \: (\neq 0)) \: \id{shape} \\
  & \quad \Let \: \id{koffsets} = \Map \: (\lambda i \to \id{offsets}[i]) \: \id{hashes}  \\
  & \quad \Let \: \id{okeys} =  \Zip \: \id{koffsets} \: \id{keys} \\
  & \quad \Let \: \id{ishape} = \\
  & \qquad (\Filter \: ((\neq 0) \circ \Snd) \circ \Zip \: (\Iota \: n)) \: \id{shape} \\
  & \quad \Let \: (\id{is}, \id{consts}) = (\Unzip \circ \kw{segmake}'_2 \: \id{okeys}) \: \id{ishape} \\
  & \quad \In \: \Scatter \: (\Rep \: n \: (\Random \: ())) \: \id{is} \: \id{consts}
\end{align*}
Now the trick to avoid sorting is to associate every key with the index of the
subarray it belongs to.
\begin{align*}
  & \mathbf{def} \: \kw{seghashes} \: (\id{shape}: [m]\Int) \: (\id{consts}: [m][c]\Int) \:  \\
  & \qquad (\id{keys}: [n](\Int, \alpha)) = \\
  & \quad \Let \: \id{offsets} = (\Presum \circ \Map (\lambda s \to s^2)) \: \id{shape} \\
  & \quad \In \: \Map \: (\lambda (\id{o}, \id{key}) \to \\
  & \quad \quad \Let \: (\id{o'}, \id{cs}, \id{s}) = (\id{offsets}[\id{o}], \id{consts}[\id{o}], \id{shape}[\id{o}]) \\
  & \quad \quad \In \: \id{o'} + (\Hash \: \id{cs} \: \id{key}) \: \Mod \: s^2) \: \id{okeys}
\end{align*}
This in turn means that the hashes can be computed by using the index of which
subarray they belong to instead of computing $\id{ii}_2$ and relying on the keys
being in order.
\begin{align*}
  & \mathbf{def} \: \kw{segmake}'_2 \: (\id{keys}: [n](\Int, \alpha)) \: (\id{ishape}: [m](\Int, \Int)) = \\
  & \quad \If \: n = 0 \: \Then \: [] \: \Else \\
  & \quad \Let \: \id{shape} = \Map \: \Snd \: \id{ishape} \\
  & \quad \Let \: \id{consts} = \kw{segrandom} \: \id{shape} \\
  & \quad \Let \: \id{hashes} = \kw{seghashes} \: \id{shape} \: \id{consts} \: \id{keys} \\
  & \quad \Let \: \id{collisions} = \kw{segcollisions} \: \id{shape} \: \id{hashes} \\
  & \quad \Let \: (\id{ishape'}, \id{iconsts'}) = \\
  & \qquad \kw{segresult} \: \id{ishape} \: \id{consts} \: \id{collisions} \\
  & \quad \Let \: \id{keys'} = \Filter \: (\lambda (\id{o}, \id{key}) \to \id{collisions}[\id{o}]) \: \id{keys} \\
  & \quad \Let \: \id{offsets} = \Presum \: \id{collisions} \\
  & \quad \Let \: \id{keys}'' = \Map \: (\lambda (\id{o}, \id{key}) \to (\id{offsets}[\id{o}], \id{key})) \: \id{keys'} \\
  & \quad \In \: \id{iconsts'} \concat \kw{segmake}'_2 \: \id{keys''} \: \id{ishape'}
\end{align*}
It is then also possible to remove keys by using the index into its subarray by
checking if the subarray has no collisions according to the collision array.
Another problem arises now which is after removing the size of subarrays in
\textit{shape} the offset of a given key will no longer have the correct index.
This can be resolved by taking a prefix-sum of the \textit{collisions} array and
update the index of the key to be the offset of the new subarray it belongs to
after filtering.

\subsection{Analysis}
Once again almost every line of of code in $\kw{make}_1$ is $O(n)$ since every
line of code does at most $O(n)$ work besides $\kw{segmake_2}$.
It ends up doing the expected work of $O(n)$ since it works on every subarray of
size $n_i^2$ and it is expected to work on each subarray at most $2$ times.
And all the work done internally by $\kw{segmake}_2$ is using functions which
does at most the expected work of $O(n)$.
So in total the algorithm has the expected work of $O(n)$ hence it is work
efficient.

For the span of the algorithm we see a segmented-or over the flattened array so
which has an expected span of $O(\log n)$.
No other operation has a worse span, this means that the expected span of the
algorithm is $O(\log n)$.

\section{Interface}
\label{sec:interface}

In this section we will describe our abstract programming interface for
key-value maps in a functional array language.
The goal is to define an interface that can allow keys of any type, as long a
universal hash function and a total ordering exists for the type, and values of
any type.
We will use the concrete syntax of Futhark, including its ML-derived module
system~\cite{Elsman:2018:SIH:3243631.3236792}, but we will assume no familiarity
and describe the concepts as necessary.
For a primer on the Standard ML module system, see~\cite{tofte1996essentials},
although note that Futhark diverges somewhat in syntax and nomenclature.
The interface is condensed compared to our actual implementation, in order to
avoid cluttering the central ideas.

\subsection{Irregularly Sized Keys}

One major technical challenge is how to handle keys of irregular size (e.g.,
strings), which is not directly supported in Futhark.
The problem is that Futhark does not support \emph{irregular arrays} (sometimes
called \emph{jagged arrays}), meaning arrays where elements can have different
sizes.
When strings are modeled as arrays of characters, this prevents the key array
from being representable, unless all the keys happen to be strings of the same
length---which is an unlikely case.
While our implementation is in Futhark, the requirement for regular arrays is a
common limitation among functional array languages, such as
Accelerate~\cite{chakravarty2011accelerating} and SaC~\cite{SCHOLZ_2003}, and is
ultimately rooted in fundamental issues the efficient representation of arrays
in memory.
The NESL language~\cite{blelloch1995nesl} does support irregular arrays (through
a flattening transformation), but the resulting representation, when done
completely automatically, tends to be rather inefficient.
APL~\cite{iverson1962programming} supports a form of irregular arrays by
``boxing'' the elements, essentially turning them into scalars, but the
resulting in-memory representation becomes an array of pointers, which can lead
to poor locality.
Most of these languages are targeted at numerical problems, where the inability
to represent irregular arrays is less of a problem, but it does pose a challenge
when we wish to represent something as basic as an array of strings.

One common solution, and the one we will use, is to represent a string as a pair
of an \emph{offset} and a \emph{length} into some other array of characters,
which we call the \emph{context} (the choice of this term will be made clear
shortly).
The idea is that the context essentially consists of the concatenation of all
strings of interest, and we slice it as necessary to obtain substrings.
The context is not a \emph{string pool}, as we do not assume that a string is
\emph{uniquely} determined by its offset and length - it may well be that two
strings with different offsets actually correspond to sequences of equivalent
characters in the context.

In some languages, it would be possible for a string contains a reference to the
context from which its characters are sliced.
This is not viable in Futhark, or most similar value-oriented array languages,
as they do not support references or other such pointer-like types, and a
reference to the context would imply a copy.
Instead, the coupling between a string and its context is implicit, and for all
functions defined on strings---such as ordering or hashing---the caller must
pass the context as well.
Passing the wrong context by accident is indeed a potential source of errors,
and one that Futhark's type system is not sufficiently strong to avoid.
In our implementation of key-value maps, we mandate that all keys stored in the
map must use the same context (which we will also store in the map itself), but
when querying membership, the ``needle'' need not be.

\subsection{The Map Interface}
\label{sec:map}

The abstract interface for our key-value maps, expressed as a module type
(\emph{signature} in Standard ML), is shown in \cref{fig:map-module-type}.
A module type specifies types, values, and functions that must be provided by an
implementation of the interface.
In the module type, all types are abstract, but an implementation will often
\emph{refine} some of these types to be concrete.

The \texttt{map} module type comprises three abstract types: \texttt{key}
denotes type of keys (which an implementation will refine to be a specific type,
e.g., \texttt{i32}), the \texttt{ctx} is the context used for comparison and hash
functions, and \texttt{map [n] v} is the type of a map containing \texttt{n}
key-value mappings to values of type \texttt{v}.
The \texttt{[n]} is a \emph{size parameter}~\cite{10.1145/3609024.3609412},
which is used to document how the sizes of function arguments relate to function
outputs, although it is only of minor importance to the present work.
The \texttt{map} type is declared with the keyword \lstinline{type~}, which is
Futhark notation for an abstract type that may internally contain arrays of
unspecified size---this implies that we may not construct arrays with elements
of type \texttt{map}, as this would essentially result in irregular arrays.

The \texttt{map} module type also specifies three API functions:
\texttt{from\_array} constructs a map from from an array of key/value-pairs,
\texttt{from\_array\_nodup} does the same but assumes that no duplicate keys
exist (saving us deduplication during construction), and \texttt{lookup}
retrieves the value corresponding to a specified key, using an option type to
handle the case where no such key exists in the map.

\begin{figure}
\begin{lstlisting}[language=Futhark]
module type map = {
  type key
  type ctx
  type~ map [n] 'v

  val from_array [u] 'v :
    ctx -> [u](key, v) -> ?[n].map [n] v

  val from_array_nodup [n] 'v :
    ctx -> [n](key, v) -> map [n] v

  val lookup [n] 'v :
    ctx -> key -> map [n] v -> opt v
}
\end{lstlisting}
  \label{fig:map-module-type}
  \caption{The \texttt{map} module type, which specifies the abstract types and
    functions of the key-value API. This is a minimal API that elides many
    convenient functions that are present in our full implementation, but are
    not interesting from a data-parallel perspective.}
\end{figure}

\subsection{The Key Interface}
\label{sec:key}

The \texttt{map} interface can be implemented in many ways, but an
implementation is always specialised for a given \texttt{key} type, as the
\texttt{map} type itself is not polymorphic in the keys, only in the values.
Clearly it is not desirable to reimplement the interface from scratch whenever
we wish to use a new key type.
We solve this with \emph{parameterised modules}
(\emph{functors} in Standard ML), which can be seen as module-level functions
that are applied at at compile time to form new modules.
Essentially, we write a parameterised module that, given a module that provides
a key type and necessary operations on keys, constructs a module that implements
the \texttt{map} module type for the supplied key type.

We describe our requirements for keys in a module type \texttt{key}, shown in
\cref{fig:key-module-type}.
Three types are specified: \texttt{ctx} is the context as discussed above,
\texttt{key} is the key type itself, and finally \texttt{uint} is the type of
hashes.
We also require three value definitions: \texttt{m} is the number of constants
passed to the hash function as discussed in \cref{seq:prerequisites},
\texttt{hash} is a function for hashing a value of type \texttt{key}, and
\texttt{(<=)} is the comparison operator.
Both \texttt{hash} and \texttt{(<=)} are provided a context. For the latter, two
contexts are in fact used: one for each value, which is crucial when performing
lookups, as the ``needle'' key is likely to use a different context than
``haystack'' keys.

\Cref{fig:key-module-type} also shows a binding of a module \texttt{i32key} that
implements the \texttt{key} module type where all types have been made concrete.
The implementation has been elided, but it is based on a universal hash
function. Note that the context type has been refined to the unit type, as no
context is needed for integer keys.

\begin{figure}
  \centering
\begin{lstlisting}[language=Futhark]
module type key = {
  type ctx
  type key
  type uint

  val m : i64
  val hash : ctx -> [m]uint -> key -> uint
  val (<=) : (ctx, key) -> (ctx, key) -> bool
}

module  u8key : key with ctx = ()
                    with key = u8
                    with uint = u64
  = ...

module i32key : key with ctx = ()
                    with key = i32
                    with uint = u64
  = ...
\end{lstlisting}
  \caption{The \texttt{key} module type, which must be implemented for any type
    used as a key, as well as a binding of modules \texttt{u8key} and
    \texttt{i32key} that declare they implements a refinement of \texttt{key}
    with all abstract types made concrete.}
  \label{fig:key-module-type}
\end{figure}

\subsubsection{Array Slices as Keys}

We can now define a representation of array slices that implement the
\texttt{key} module type.
In \cref{fig:slice-module} we first define a module type \texttt{slice} that
specifies the rather simple interface for array slices: constructing slices from
offset and length, decomposing a slice into offset and length, and finally
applying the slice to a concrete array, returning another array of some unknown
size \texttt{k}.
The type \texttt{slice a} denotes a slice of an array with element type
\texttt{a}---in practice, \texttt{a} is a phantom type.
The implementation of the module type (as a module also called \texttt{slice})
is straightforward and also shown in \cref{fig:slice-module}.

The \texttt{mk\_slice\_key} parameterised module accepts a \texttt{key} module
\texttt{E} and produces a \texttt{key} module for keys of type \texttt{slice
  E.key}, and context \texttt{[]E.key}. Importantly, the element type itself
must not require any context. We elide the implementation for brevity, but note
that the implementation of a universal hash function for sequences is somewhat
complicated, although unrelated to data-parallelism and outside the scope of
this paper.

\begin{figure}
\begin{lstlisting}[language=Futhark]
module type slice = {
  type slice 'a
  val mk 'a : (i: i64) -> (n: i64) -> slice a
  val unmk 'a : slice a -> (i64, i64)
  val get [n] 'a : slice a -> [n]a -> ?[k].[k]a
}

module slice : slice = {
  type slice 'a = {i: i64, n: i64}
  def mk i n = {i, n}
  def unmk {i, n} = (i, n)
  def get {i, n} xs = xs[i:i + n]
}

module mk_slice_key (E: key with ctx = ()
                            with uint = u64)
  : key with ctx = []E.key
        with key = slice.slice E.key
        with uint = u64 = {
  ...
}
\end{lstlisting}
  \caption{The \texttt{slice} module type and module, as well as a parameterised
    module \texttt{mk\_slice\_key} that constructs a key module for slices of
    arrays of some type, given a module that implements \texttt{key} for the
    element type, with a unit context.}
  \label{fig:slice-module}
\end{figure}

\subsection{Constructing Maps}
\label{sec:constructing-maps}

Now that we have established the \texttt{key} and \texttt{map} abstractions, we
can defined parameterised modules that produce implementations of \texttt{map}.
\Cref{fig:mkarraymap} sketches an implementation that represents the mapping as
a sorted array of keys and a corresponding array of values.
Construction then requires sorting (and possibly deduplicating) the provided
key/value-pairs, and \texttt{lookup} is implemented as a binary search.
This is a fairly simple implementation, assuming that one has access to an
efficient sorting function.
Note that with our definition of the \texttt{key} module type, radix sorting is
not possible (because we provide no way to decompose into bits), so instead our
implementation uses merge sort.
As a locality optimisation, instead of storing the keys in sorted order, we can
use an Eytzinger layout that essentially stores each level of the corresponding
binary tree that is implicitly formed when performing the binary
search~\cite{10.1145/512274.3734138}.

We show the implementation of \texttt{lookup} in order to demonstrate how
contexts are passed around.
We assume the existence of a function \texttt{binary\_search : (v -> v -> bool)
  -> []v -> v -> i64} that given a comparison operator performs a binary search
for the index of a provided needle in a sorted array, and returns \texttt{-1} in
case the needle is not found.
It is crucial that \texttt{lookup\_index} provides the right context to the
\texttt{K.<=} operator.
In particular, the \texttt{x} parameter is drawn from the arraymap itself, while
\texttt{y} is the needle.
This means the former must be paired with the the \texttt{ctx} from the map, and
the latter with the \texttt{ctx} provided by the caller of \texttt{lookup}.
Unfortunately these are both of type \texttt{ctx}, so mixing them up will not
cause a type error---we will instead obtain wrong results at runtime, and
somewhat insidiously, the issue will only occur for keys that actually make use
of the contexts, and not for the common case of, for instance, integer keys.

\begin{figure}
\begin{lstlisting}[language=Futhark]
module mk_arraymap (K: key)
  : map with key = K.key
        with ctx = K.ctx = {
  type key = K.key
  type~ ctx = K.ctx

  type~ map [n] 'v =
    { ctx: ctx
    , keys: [n]key
    , vals: [n]v
    }

  def lookup_index [n] 'v
      (ctx: ctx) (k: key) (m: map [n] v) =
    binary_search
      (\x y -> (m.ctx, x) K.<= (ctx, y))
      m.keys
      k

  def lookup [n] 'v
      (ctx: ctx) (k: key) (m: map [n] v) =
    match lookup_index ctx k m
    case -1 -> #none
    case i -> #some m.vals[i]

  ...
}
\end{lstlisting}
  \caption{Outline of the \texttt{mk\_arraymap} parameterised module, which
    given an implementation of \texttt{key} constructs an implementation of
    \texttt{map}. The value definitions have been elided for brevity.}
  \label{fig:mkarraymap}
\end{figure}

Of course, given the topic of the present paper, an implementation based on
binary search is not satisfactory.
We therefore also define \texttt{mk\_hashmap}, shown in \cref{fig:mkhashmap},
which is based on two-level hash maps following the approach of
\cref{sec:two-level-hashmaps}.
Note the complete equivalence with the interface in \cref{fig:mkarraymap}.

\begin{figure}
\begin{lstlisting}[language=Futhark]
module mk_hashmap (K: key)
  : map with key = K.key
        with ctx = K.ctx = {
  type~ map [n] 'v =
      ?[f][m].
      { ctx: K.ctx
      , key_values: [n](K.key, v)
      , offsets: [f]i64
      , lookup_keys: [f]K.key
      , level_one_consts: [K.c]uint
      , level_two_consts: [m][K.c]uint
      , level_two: [n][3]i64
      , rng: rng
      }

  ...
}
\end{lstlisting}
  \caption{Outline of the \texttt{mk\_hashmap} parameterised module, which given
    an implementation of \texttt{key} constructs an implementation of
    \texttt{map} based on two-level hash maps as discussed in
    \cref{sec:two-level-hashmaps}. We assume the presence of a type \texttt{rng}
    that is a the state of a random number generator. The sizes \texttt{f} and
    \texttt{m} are internally bound and not visible in the external type.}
  \label{fig:mkhashmap}
\end{figure}

While the definition of these modules is somewhat intricate, applying the
building blocks is straightforward, as shown in \cref{fig:map-modules}.
The resulting modules can be used without knowledge of module-level programming,
and it is straightforward to provide predefined modules for common key types in
a library.
Some of the generality in the underlying design does leak through in undesirable
ways.
For example, the type of \texttt{i32hashmap.lookup} is \texttt{() -> i32 ->
  i32hashmap.map [n] v -> opt v}.
The \texttt{()} parameter is the context, which is the unit type for this
module.
It is not difficult to write wrapper definitions that remove the need for
passing a unit value to \texttt{lookup}, but it is somewhat tedious, and
requires replicating the entire \texttt{map} module type---which is
significantly larger in a practically useful library than the subset we show in
\cref{fig:map-module-type}, although the cost is entirely on the implementer of
the module, not on the user.

\begin{figure}

\begin{lstlisting}[language=Futhark]
module i32arraymap = mk_arraymap i32key
module i32hashmap = mk_hashmap i32key
module strkey = mk_slice_key u8key
module stringhashmap = mk_hashmap strkey
\end{lstlisting}

  \caption{Modules for various implementations of the \texttt{map} module type,
    where we consider strings to be simply arrays of unsigned 8-bit integers.}
  \label{fig:map-modules}
\end{figure}

\section{Benchmarks}
\label{sec:benchmarks}

We have implemented two-level hash maps in Futhark, which is a functional array
language that can generate usually quite efficient GPU
code~\cite{futhark/sc22mem,Henriksen:2019:IFN:3293883.3295707,henriksen:phdthesis}.
To evaluate the performance of our hash maps, we have constructed a set of
benchmarks.
They are all based around inserting $n$ keys of two types (64-bit integers and
strings of 5--25 characters), with the values always being $32$-bit integers.
The keys are uniformly distributed and generated such that there are no
duplicates.
For various implementations of key-value maps, we benchmark the time it takes to
perform the following operations:

\begin{enumerate}
\item \emph{Construction} of the map (e.g., building a hash map or sorting an array).
\item \emph{Lookup} of the value associated with every key in the map.
\item \emph{Membership} testing of every key---this differs from lookup in
  that the value is not retrieved.
\end{enumerate}

We benchmark the following implementations:

\begin{enumerate}
\item Futhark, using two-level hash maps.
\item Futhark, using binary search.
\item Futhark, using binary search with an Eytzinger layout.
\item cuCollections~\cite{cuCollections}, a implementation of hash maps in CUDA
  C++, which is to our knowledge the state of the art in hash maps on GPUs. The
  static hash map in cuCollections is not based on two-level hash maps, but
  rather uses open addressing with linear probing.
\end{enumerate}
We run our benchmarks on an NVIDIA A100 GPU, using Futhark's CUDA backend, and
without disabling bounds checking (which adds about 5\%
overhead~\cite{hlpp20-futhark}). The results are shown in \cref{tab:benchmarks}.
The most obvious result is that cuCollections is vastly faster than Futhark in
all cases; ranging from nearly $10\times{}$ speedup in the construction of the hash
table, to $1.8\times{}$ for membership testing.
We will return to this in \cref{sec:whatistobedone}.
The results suggest that even in a data-parallel language and for integral keys,
hash maps bringmeaningful speedup compared to comparison/tree-based techniques.
Because we assume unique keys, constructing the hash maps is not even slower than
sorting key/value arrays, although this not be the case if we had to perform
deduplication of keys.

\begin{table*}
  \centering
  \begin{tabular}{l|rrr|rrr}
    & \multicolumn{3}{c|}{\textbf{64-bit integer keys} ($n=10^{7}$)}
    & \multicolumn{3}{c}{\textbf{String keys} ($n=10^{7}$)} \\
    & \textit{Construction} & \textit{Lookup} & \textit{Membership}
    & \textit{Construction} & \textit{Lookup} & \textit{Membership} \\\midrule
    Futhark (hash maps) & 18.3 & 3.3 & 1.6 & 33.2 & 4.3 & 2.8 \\
    Futhark (binary search) & 40.9 & 6.2 & 5.8 & 83.0 & 5.7 & 5.8 \\
    Futhark (Eytzinger) & 42.3 & 4.3 & 2.4 & 85.3 & 5.3 & 5.3 \\
    cuCollections & $2.7$ & $1.1$ & $0.9$ & $2.7$ & $1.3$ & $1.2$  \\
  \end{tabular}
  \caption{Benchmark results. All runtimes are in milliseconds.}
  \label{tab:benchmarks}
\end{table*}

\subsection{Future Work}
\label{sec:whatistobedone}

It is clear that the hash maps implemented in cuCollections significantly
outperform the ones we have implemented in Futhark.
This is not surprising---cuCollections is developed by GPU experts with a track
record of research into GPU hash
maps~\cite{cudahash,DBLP:conf/hipc/JungerKM0XLS20}, and using a low-level
language that allows tuning and exploitation of hardware properties.
It is unlikely that an implementation in a high level and hardware-agnostic
language can match this.
Yet it may still be possible to narrow the gap, and this may improve high level
implementations of \emph{other} algorithms, which may not already been carefully
implemented in low level languages by experts.

The fastest GPU hash map implementations are open addressing hash maps,
implemented as concurrent data structures, with multiple threads simultaneously
inserting
elements~\cite{DBLP:conf/hipc/JungerKM0XLS20,DBLP:conf/ipps/Junger0S18}, with
synchronisation done through atomic operations.
While open addressing hash maps can be implemented in Futhark (and we have
done so as an experiment), we find that the efficient implementations that are
possible in CUDA cannot be expressed using the available data-parallel
vocabulary, and they perform significantly worse than two-level hash maps.

The core problem is that although hash maps provide a deterministic interface,
their efficient (concurrent) implementation is \emph{internally}
nondeterministic---for instance, when two threads compete for the same slots in
an open addressing hash map, it is nondeterministic who gets which slot,
although the choice is not externally visible.
When implementing a data structure in a deterministic data-parallel language,
such as Futhark and most other array languages, we are not at liberty to locally
exploit nondeterminism and impurity the way we can do in CUDA.
Of course, the flip side is also that the language prevents us from accidentally
introducing nondeterminism---the race-freedom of cuCollections is due to careful
implementation and testing, rather than a guaranteed property of CUDA itself.
However, it does seem functional array languages are missing a mechanism for
locally exploiting nondeterministic operations, which can be encapsulated in
such a way that the nondeterminism can be localised, but it is not clear what
such a mechanism should look like.
The Dex language uses an effect system to allow local mutation of an array
shared between multiple iterations of a parallel loop~\cite{10.1145/3473593},
although the only effect allowed without hindering parallelism is incrementing
an array element.
This is not sufficient for our needs, and indeed the restriction to increments
is what provides determinism.

Another point of view is that the language should provide hash maps as a built-in
type, which can then be implemented with low-level code in the language
implementation itself.
The arrays in an array language are implemented this way, so it is certainly
conceivable that hash maps could be provided similarly---and some APL dialects do
indeed provide such built-in data structures.
While pragmatic, we do consider this approach less compelling from a research
perspective, and certainly less ambitious from an implementation perspective.

% Allow for atomics.
% Probably something about a warp should be better at communicating.

\section{Related Work}
There are several implementations of hash maps meant for GPUs.
The warpcore and warpdrive
libraries~\cite{DBLP:conf/hipc/JungerKM0XLS20,DBLP:conf/ipps/Junger0S18} are
written in CUDA and also has an implementation of open addressing hash maps.
They do a similar parallel probing scheme like cuCollections such that every
thread in a warp works together to find a key to achieve coalesced access
patterns.
The warpcore library also contains a bucket list hash table, where each bucket
contains a linked list of key-value pairs.

The consolidation of the aforementioned libraries are
cuCollections~\cite{cuCollections}. Which contains a fast implementation of open
addressing hash maps and is used for benchmarking in this paper.

BGHT~\cite{Awad:2023:AAI} has static hash maps which both uses bucketed cuckoo
hash tables and iceberg hashing.
These hash tables are implemented in CUDA and are designed to be used in a
concurrent setting.

The aforementioned libraries are all written in CUDA, but there also exists an
implementation of a hash map\footnote{The implementation can be found at
\url{https://www.cs.cmu.edu/afs/cs.cmu.edu/project/scandal/public/code/nesl/nesl/examples/hash-table.nesl}.}
written in the high-level functional array language
NESL~\cite{blelloch1995nesl}.
It is an open addressing hash map with quadratic probing, what they roughly do
is try to insert every key in parallel.
All keys that lead to a collision because the slot is already occupied, try
recursively to insert the keys into the next slot using quadratic probing.
And if the slot is empty then select one of the keys to insert into that slot,
the remaining are then inserted in the recursive call using the probing scheme.
Such an implementation could lead to nondeterminism depending on the
implementation but you could also use some rule to select which key to insert
into the empty slot, like we do for our open addressing hash map written in
futhark.
The reason for this not being benchmarked against in the present paper is that
no compiler is available for NESL that can generate code for our GPUs.

\section{Conclusions}
We have shown in detail how to derive a data-parallel construction of two-level hash maps by flattening a functional formulation.
The resulting algorithm is work efficient, meaning it has an expected work of
$O(n)$ and an expected span of $O(\log n)$.

We describe how such key/value data structures can be provided in a generic and
abstract way, demonstrated by an implementation in the functional array language
Futhark.
Our approach allows for irregular keys, such as strings, by representing them as
an offset and length into a context array of characters, although the interface
does allow mismanagement of said context that is not detected statically.

Our performance evaluation shows that hash maps performs well compared to a
binary search based implementations, but is much slower than state of the art
GPU hash maps implemented in CUDA.
The performance differences are partly due to the CUDA implementations being
carefully tuned to take advantage of specific hardware properties, but also due
to the CUDA implementation using a more efficient data structure---open
addressing hash tables---which are difficult to implement efficiently in our
data-parallel vocabulary.
Furthermore, our two-level hash maps uses more memory than open addressing and
will most likely lead to two cache misses per lookup.
Unlike open addressing which will most likely only lead to one cache miss per
lookup.

Finally we have briefly discussed the problems of implementing efficient
concurrent data structures in functional array languages, for which the central
challenge seems to be allowing some forms of controlled nondeterminism.

\bibliography{arxiv25-hashmaps}

\end{document}